\documentclass[12pt,times]{article}
\usepackage[top=1.5cm,bottom=2.5cm,left=1.5cm,right=1.5cm]{geometry}

\usepackage[figuresright]{rotating}
\usepackage{extarrows}
\RequirePackage{lineno}
\usepackage[colorlinks=true]{hyperref}
\usepackage{fancyhdr,amsmath,amssymb,graphicx,bm,upgreek}
\usepackage{multirow,multicol,array,color,mathrsfs,subcaption,booktabs}
\usepackage{epstopdf,caption}
\usepackage{pxfonts}
\usepackage[T1]{fontenc}
\usepackage{authblk}

\usepackage[numbers,sort&compress]{natbib}

\DeclareGraphicsExtensions{.eps,.mps,.pdf,.jpg,.png}
\graphicspath{{figure/}{../figure/}}

\title{Controlled release of entrapped nanoparticles from thermoresponsive hydrogels with tunable network characteristics}
\author{Yi Wang$^{1,2}$,
        Zhen Li$^{3}$\footnote{Corresponding author.
        \\\indent~~~Email addresses:
        \href{mailto:zli7@clemson.edu}{zli7@clemson.edu},~
        \href{mailto:jieouyang@nwpu.edu.cn}{jieouyang@nwpu.edu.cn},~
        \href{mailto:george_karniadakis@brown.edu}{george\_karniadakis@brown.edu}},
        Jie Ouyang$^{1}$~and
        George Em Karniadakis$^{2}$\\
        \small{$^{1}$Department of Applied Mathematics, Northwestern Polytechnical University, Xi'an 710129, China}\\
        \small{$^{2}$Division of Applied Mathematics, Brown University, Providence, Rhode Island, 02912, USA}\\
        \small{$^{3}$Department of Mechanical Engineering, Clemson University, Clemson, SC 29634, USA}
}

\date{}

\begin{document}
\maketitle

\vspace{-0.75cm}
\begin{abstract}
Thermoresponsive hydrogels have been studied intensively for creating smart drug carriers and controlled drug delivery. Understanding the drug release kinetics and corresponding transport mechanisms of nanoparticles (NPs) in a thermoresponsive hydrogel network is the key to the successful design of a smart drug delivery system. To investigate the anomalous NP diffusion in smart hydrogels with tunable network characteristics, we construct an energy-conserving dissipative particle dynamics model of rigid NPs entrapped in a hydrogel network in an aqueous solution, where the hydrogel network is formed by cross-linked semiflexible polymers of thermoresponsive poly({\em N}-isopropylacrylamide)~(PNIPAM). By varying the environmental temperature crossing the lower critical solution temperature of PNIPAM we can significantly change the hydrogel network characteristics.  We systematically investigate how the matrix porosity of the hydrogel and the nanoparticle size affect the NPs' diffusion process and NPs' release kinetics at different temperatures. Quantitative results on the mean-squared displacement and the van Hove displacement distributions of NPs show that all NPs entrapped in the smart hydrogels undergo subdiffusion at both low and high temperatures. For a coil state and as the system temperature approaches the critical temperature of the coil-to-globule phase transition, both the subdiffusive exponent and the diffusion coefficient of NPs increase due to the increased kinetic energy and the decreased confinement on NPs, while the transport of NPs in the hydrogels can be also enhanced by decreasing the matrix porosity of the polymer network and NPs' size. However, when the solution temperature is increased above the critical temperature, the hydrogel network collapses following the coil-to-globule transition, with the NPs tightly trapped in some local regions inside the hydrogels. Consequently, the NP diffusion coefficient can be reduced by two orders of magnitude, or the diffusion processes can even be completely stopped. These findings provide new insights for designing controlled drug release from stimuli-responsive hydrogels, including autonomously switch on/off drug release to respond to the changes of the local environment.
\end{abstract}

\section{INTRODUCTION}\vspace{-0.15cm}
Stimuli-responsive hydrogels are three-dimensional polymer networks, which are capable of absorbing and retaining vast amounts of water. The structure properties of these hydrogels can change dramatically in response to external stimuli changes~\cite{qiu2001environment,ganta2008review}, including pH~\cite{kreft2007polymer}, temperature~\cite{yeh2015mesoscale}, light intensity~\cite{timko2014near}, and magnetic/electric fields~\cite{olsson2010making}. Stimulus-sensitive hydrogels have attracted considerable attention over the past two decades because of their great potential in the fields of biology and medicine~\cite{liu2010recent,wei2017stimuli}. Examples include drug delivery~\cite{elsabahy2012design}, artificial muscles~\cite{dicker2017light}, and programmable soft micromachines~\cite{huang2016soft}. Poly($N$-isopropylacrylamide) (PNIPAM) is one of the most popular thermoresponsive polymer, which experiences a reversible phase transition around a lower critical solution temperature (LCST) of 32~$^{\circ}$C. Below the LCST the PNIPAM is hydrophilic and swells in water, while above the LCST it becomes hydrophobic and expels water resulting in the collapse of hydrogels and a sharp decrease in gel volume. The characteristic of a LCST at 32~$^{\circ}$C between the room temperature and the physiological temperature makes PNIPAM the most extensively studied temperature-sensitive polymer for biomedical applications~\cite{kang2016collapse,zha2011stimulus,shirakura2014hydrogel}.

Because of the high thermosensitivity and good biocompatibility, PNIPAM becomes especially attractive in the applications of drug delivery~\cite{roy2013new}. PNIPAM acting as the drug delivery carrier can protect encapsulated drugs until they reach the targeted sites, and then release the drugs from hydrogels on demand, thereby playing an important role in targeted and controlled drug delivery systems ~\cite{schmaljohann2006thermo,masoud2011controlled}. In practical applications of thermoresponsive hydrogels, taking the temperature as a stimulus for controlled release of drugs from hydrogels has been successfully applied in many cancer treatments~\cite{zhang2014near,liu2015temperature,cho2008therapeutic}. With an increased metabolic rate, the tumor tissue is generally at a  temperature slightly higher than the physiological temperature of healthy tissues~\cite{schmaljohann2006thermo,stefanadis2001increased}. The temperature difference between tumor and healthy tissue can be utilized by thermoresponsive hydrogels to achieve the controlled drug release, i.e., quick and efficient release of anticancer drugs to tumor tissue while little release elsewhere. The controlled drug release not only improves therapeutic efficacy but also reduces systemic side effects of anticancer drugs. Understanding the transport mechanism of drugs and nanoparticles (NPs) contained in thermoresponsive hydrogels is critical for designing controlled release of anticancer drugs. In general, diffusion is the dominant mechanism for the transport of drugs and NPs in the hydrogel networks. Investigating the NP diffusion entrapped in the thermoresponsive hydrogels and understanding how the local hydrogel structures affect the NP transport process are important for controlled release of small molecules and drugs from the hydrogels.

Although the NP diffusion in polymeric fluids has been extensively studied using theoretical~\cite{yamamoto2011theory,yamamoto2014microscopic,dong2015diffusion,cai2015hopping, yamamoto2018theory} and experimental~\cite{kohli2012diffusion,gam2011macromolecular,carroll2018diffusion,poling2015size,babaye2014mobility} approaches as well as computer simulations~\cite{kalathi2014nanoparticle,patti2014molecular,volgin2017molecular,sorichetti2018structure,liu2008molecular} during the past years, these works have mainly focused on the motion of NPs in polymer solutions or melts. Few studies have been reported on the diffusion of NPs entrapped in thermoresponsive hydrogels and there are no effective models for predicting the mobility of NPs in a polymeric network with thermally induced phase transitions. Because of a lack of effective mathematical or physical models, many experimental efforts have been made to determine the diffusion of NPs using single particle tracking (SPT)~\cite{parrish2018temperature,parrish2017network,stempfle2014anomalous} or measuring the uptake and release of drugs into thermo-sensitive hydrogels~\cite{constantin2011lower,seden2007investigation}. However, in these works, experimental characterization can be laborious and expensive, or only a few factors were considered. Understanding and quantifying the {\em anomalous NP diffusion process} encountered in thermoresponsive hydrogels requires a systematic study on the impact of tunable network characteristics of hydrogels and NP size on the NPs' diffusion process as well as the NPs' release kinetics at different temperatures.

In the present work we systematically study the anomalous diffusion process of NPs contained in thermoresponsive hydrogels using energy-conserving dissipative particle dynamics (eDPD) simulations. The eDPD model is a particle-based mesoscopic simulation method, which was developed as an extension of the classical DPD method~\cite{avalos1997dissipative, espanol1997dissipative}. Because the classical DPD model was developed based on equilibrium thermodynamics~\cite{espanol1995statistical} and is limited to modeling isothermal systems, the eDPD model incorporates the internal energy as an extra attribute of each DPD particle for considering the mesoscopic energy equation. Therefore, eDPD not only preserves the conservation of system energy, but also correctly reproduces the thermodynamic properties of fluids~\cite{li2014energy}, and has been successfully applied to many interesting non-isothermal processes, including natural convection~\cite{cao2013energy}, thermoresponsive microgels~\cite{li2015mesoscale} and thermoresponsive micelles and vesicles~\cite{tang2016non}. The capability of eDPD method on modeling non-isothermal processes provides us a powerful tool to study the controlled release of entrapped nanoparticles from thermoresponsive hydrogels.

The reminder of the paper is organized as follows: In section~\ref{sec:2}, we present the details of the eDPD model and its parameterization for modeling the NP diffusion inside the thermoresponsive hydrogels. In section~\ref{sec:3}, we systematically study and quantify the diffusion process and the controlled release of NPs entrapped in the thermoresponsive hydrogels, with a special focus on the effects of polymer concentration, NP size, and temperature. Finally, we conclude with a brief summary and discussion in section~\ref{sec:4}.

\section{METHOD}\label{sec:2}\vspace{-0.15cm}
\subsection{Energy conserving DPD model (eDPD)}
Similar to the classical DPD method, an eDPD system consists of many coarse-grained particles representing collective dynamics of a group of actual molecules. By introducing the internal energy to each DPD particle in addition to other quantities, such as position and momentum~\cite{espanol1997dissipative}, the momentum and energy equations of an eDPD particle $i$ are given as follows~\cite{li2014energy,abu2011energy}:
\begin{align}
    \frac{\mathrm{d}\mathbf{r}_i}{\mathrm{d}t} &= \mathbf{v}_i, \\
    m_i\frac{\mathrm{d}\mathbf{v}_i}{\mathrm{d}t} &= \mathbf{F}_i = \sum_{j \neq i}(\mathbf{F}_{ij}^C + \mathbf{F}_{ij}^D + \mathbf{F}_{ij}^R), \\
    C_v\frac{\mathrm{d}T_i}{\mathrm{d}t} &= Q_i = \sum_{j \neq i}(Q_{ij}^C + Q_{ij}^V + Q_{ij}^R),
\end{align}
where $t$ is time, $\mathbf{r}_i$, $\mathbf{v}_i$ and $\mathbf{F}_i$ represent position, velocity and force vectors, and $m_i$, $C_v$, $T_i$ and $Q_i$ are mass, thermal capacity, temperature, and heat flux of the eDPD particle $i$, respectively.

The total force $\mathbf{F}_i$ imposed on the eDPD particle $i$ has three pairwise additive components, i.e., the conservative force $\mathbf{F}_{ij}^C$, the dissipative force $\mathbf{F}_{ij}^D$ and the random force $\mathbf{F}_{ij}^R$. They are expressed as
\begin{align}
    \mathbf{F}_{ij}^C &= a_{ij}(T_{ij}) \omega_C(r_{ij}) \mathbf{e}_{ij}, \\
    \mathbf{F}_{ij}^D &= -\gamma_{ij} \omega_D(r_{ij}) (\mathbf{e}_{ij} \cdot \mathbf{v}_{ij}) \mathbf{e}_{ij}, \\
    \mathbf{F}_{ij}^R &= \sigma_{ij} \omega_R(r_{ij}) \xi_{ij} \Delta{t}^{-1/2} \mathbf{e}_{ij},
\end{align}
where $r_{ij} = |\mathbf{r}_{ij}| = |\mathbf{r}_i - \mathbf{r}_j|$ is the distance between particles $i$ and $j$, and $\mathbf{e}_{ij} = \mathbf{r}_{ij}/r_{ij}$ is the unit vector from particle $j$ to $i$. $\mathbf{v}_{ij} = \mathbf{v}_i - \mathbf{v}_j$ is the relative velocity and $\Delta{t}$ is the time step. Here, $a_{ij}(T_{ij})$ is the temperature-dependent conservative force coefficient, and $\gamma_{ij}$ and $\sigma_{ij}$ are the coefficients of dissipative and random force. $T_{ij} = (T_i+T_j)/2$ is the pairwise local temperature between particles $i$ and $j$.

The heat flux from a neighboring particle $j$ to a particle $i$ includes the collision heat flux $Q_{ij}^C$, viscous heat flux $Q_{ij}^V$ and random heat flux $Q_{ij}^R$, which are given by~\cite{li2014energy,abu2011energy}:
\begin{align}
    Q_{ij}^C &= k_{ij} \omega_{CT}(r_{ij}) \biggl(\frac{1}{T_i} - \frac{1}{T_j}\biggr), \\
    Q_{ij}^V &= \frac{1}{2C_v} \Biggl\{\omega_D(r_{ij}) \Biggl[\gamma_{ij}(\mathbf{e}_{ij} \cdot                    \mathbf{v}_{ij})^2 - \frac{\sigma_{ij}^2}{m_{i}} \Biggr] - \sigma_{ij} \omega_R(r_{ij})                        (\mathbf{e}_{ij} \cdot \mathbf{v}_{ij}) \zeta_{ij} \Biggr\}, \\
    Q_{ij}^R &= \beta_{ij} \omega_{RT}(r_{ij}) \Delta{t}^{-1/2} \zeta_{ij}^e,
\end{align}
where $k_{ij}$ and $\beta_{ij}$ are the coefficients of the collisional and random heat fluxes. The parameter $k_{ij}$ is given by $k_{ij} = C_v^2 \kappa (T_i + T_j)^2/4k_{\mathrm{B}}$, in which $\kappa$ is known in the literature as heat friction coefficient~\cite{ripoll1998dissipative,abu2010natural} and $k_{\mathrm{B}}$ is the Boltzmann constant. $\omega_C(r_{ij})$, $\omega_D(r_{ij})$, $\omega_R(r_{ij})$, $\omega_{CT}(r_{ij})$, $\omega_{RT}(r_{ij})$ are the weight functions of $\mathbf{F}_{ij}^C$, $\mathbf{F}_{ij}^D$, $\mathbf{F}_{ij}^R$, $Q_{ij}^C$ and $Q_{ij}^R$, respectively. $\xi_{ij}$ and $\zeta_{ij}$ are symmetric Gaussian random variables with zero mean and unit variance. To satisfy the fluctuation-dissipation theorem, the dissipative and random force parameters should be coupled via $\sigma_{ij}^2 = 4\gamma_{ij}k_{\mathrm{B}}T_iT_j/(T_i+T_j)$ and $\omega_D(r_{ij}) = \omega_R^2(r_{ij})$, and the collisional and random heat flux parameters should be coupled via $\beta_{ij}^2 = 2k_{\mathrm{B}} k_{ij}$ and $\omega_{CT}(r_{ij}) = \omega_{RT}^2(r_{ij})$. A common choice of the weight functions for forces and heat fluxes is~\cite{li2014energy,li2015mesoscale,tang2016non,abu2011energy}
\begin{align}
    \omega_C(r_{ij}) &= \omega_R(r_{ij}) = \omega_{RT}(r_{ij}) = 1 -\frac{r_{ij}}{r_c}, \\
    \omega_D(r_{ij}) &= \omega_{CT}(r_{ij}) = \biggl(1 - \frac{r_{ij}}{r_c}\biggr)^2,
\end{align}
where $r_c$ is the cutoff radius beyond which the weight functions are zero.

\subsection{Simulation system and parameterization}
The simulation system consists of three components, i.e., polymer, solvent and NP. In the eDPD simulations, the polymer chains are usually represented by the bead-spring model. In this study, we consider linear polymer chains and adopt the Hookean spring model to describe the spring interactions between connected DPD particles in polymer chains. The elastic spring force is given as
\begin{equation}
    \mathbf{F}_{ij}^S = k_s(1-r_{ij}/r_s)\mathbf{e}_{ij},
\end{equation}
where $k_s = 200$ is the spring constant, and $r_s = 0.4r_c$ is the equilibrium bond length between two connected particles. Cross-links between different polymer chains are also modeled by the Hookean spring model~\cite{li2015mesoscale} with $k_s = 200$ and $r_s = 0.4r_c$. In the simulations, we consider hydrogels of three different sizes, which contain 1536, 2048 and 2560 cross-linked polymer chains with 50 eDPD particles per chain, respectively.

To model the phase transition of thermoresponsive hydrogels induced by temperature variations, the excess repulsion $\Delta a(T) = a_{sp}(T) - a_{ss}(T)$, in which $a_{sp}(T)$ and $a_{ss}(T)$ represent the repulsive coefficients of solvent-polymer and solvent-solvent respectively, is set to be a sigmoid function of the temperature~\cite{li2015mesoscale,tang2016non}:
\begin{equation}
    \Delta a(T) = \frac{\Delta A}{1.0 + \exp\bigl[-w \cdot (T - T_c) \bigr]},
\end{equation}
where a sharp change by $\Delta A$ occurs at $T = T_c$ (critical phase transition temperature), and the sharpness of the excess repulsion is determined by the parameter $w$. For the same type of particles, the conservative force coefficients are taken as $a_{ss}(T) = a_{pp}(T) = a_{nn}(T) = 75k_{\mathrm{B}}T / \rho$, where $s$, $p$, $n$, and $\rho$ denote solvent, polymer, NP, and the number density of DPD particle, respectively. Therefore, the temperature-dependent repulsive coefficient between polymer and solvent is
\begin{equation}
    a_{sp}(T) = \frac{75k_{\mathrm{B}}T}{\rho} + \frac{\Delta A}{1.0 + \exp\bigl[-w \cdot (T - T_c) \bigr]}.
\end{equation}
In the simulations, we set $\Delta A = 50$ and $w = 300$ so that the hydrogels are hydrophilic at low temperatures ($T < T_c$) and hydrophobic at high temperatures ($T > T_c$). The parameters used in our simulations are listed in Table 1.

\begin{table}[htbp]
    \caption{Parameters for simulations of the diffusion of NPs contained in thermoresponsive hydrogels. The symbols $s$, $p$ and $n$ represent solvent, polymer and NP, respectively.}
    \label{tab:my_label}
    \centering
    \begin{tabular}{ccc}
    \toprule
        Name                                   & Symbol          & Value\\
    \midrule
        Number density                         & $\rho$          & 4 \\
        Cutoff radius                          & $r_c$           & 1.0 \\
        Critical temperature                   & $T_c$           & 1.0 \\
        Temperature of system                  & $k_BT$          & 0.94$T_c$-1.06$T_c$ \\
        Time step                              & $\Delta t$      & 0.01 \\
        \multirow{2}*{Repulsive coefficient}   & $a_{ss}, a_{pp}, a_{nn}, a_{sn}, a_{pn}$ & $18.75k_{\mathrm{B}}T$ \\
                                               & $a_{sp}$        & $18.75k_{\mathrm{B}}T + \frac{50}{1.0+\exp[-300(T-T_c)]}$ \\
        Dissipative force coefficient          & $\gamma_{ij}$   & 4.5 \\
        Random force coefficient               & $\sigma_{ij}$   & $\sqrt{18k_{\mathrm{B}}T_iT_j/(T_i+T_j)}$ \\
        Thermal capacity                       & $C_v$           & $1.0 \times 10^5$ \\
        Heat friction coefficient              & $\kappa$        & $1.0 \times 10^{-5}$ \\
    \bottomrule
    \end{tabular}
\end{table}

To conveniently perform the eDPD simulations, dimensionless variables are introduced to nondimensionalize the eDPD system. Specifically, we choose three basic physical quantities, i.e., temperature $[T] = 300$ K, length $[L] = 10$ nm and mass $[m] = 2.50 \times 10^{-22}$ kg as the reference scale. Then, all other reference quantities can be derived from the above three basic quantities. The measured diffusivity of NPs with a radius of $R = 1$ (DPD units) in water at $T = 300$ K is $D_{\mathrm {NP}} = 7.26 \times 10^{-2}$ (DPD units). According to the Stokes-Einstein relation, the diffusion coefficient of NPs with a radius of $R = 10$ nm in water at 300 K is $D_{\mathrm {NP}}^{\mathrm W} = 2.58 \times 10^{-11}$ $\mathrm{m^2/s}$. By connecting this to the measured diffusivity, the time scale can be determined as
\begin{equation}
    [t] = \frac{D_{\mathrm {NP}}[L]^2}{D_{\mathrm{NP}}^{\mathrm W}} \approx 2.81 \times 10^{-7} \mathrm{s}.
\end{equation}

\begin{figure}[t]
\centering
\includegraphics[width=0.4\textwidth]{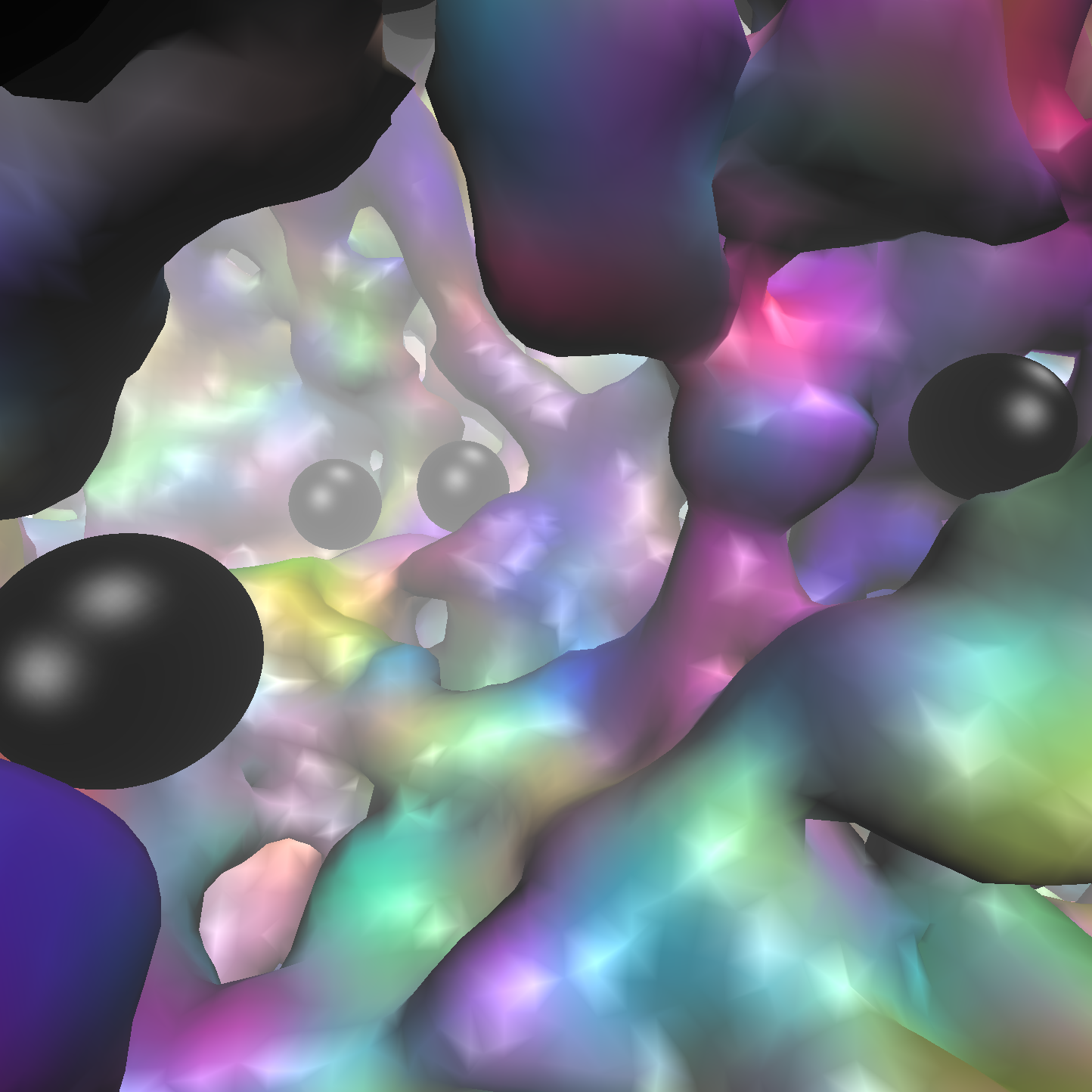}
\caption{A snapshot of the eDPD system, showing many nanoparticles (black color) moving in the interconnected porous network of thermoresponsive polymers. Solvent particles are not displayed.}
\label{fig:snapshot}
\end{figure}

In the present study, all the eDPD simulations are performed using the LAMMPS package~\cite{plimpton1995fast} and a modified velocity-Verlet algorithm~\cite{groot1997dissipative} is employed for numerically integrating the eDPD equations with the time step $\Delta t = 0.01[t]$. In our simulations, three eDPD systems containing different sizes of hydrogels are constructed in a computational domain of 400 nm $\times$ 400 nm $\times$ 400 nm. Periodic boundary conditions are applied in all three dimensions. The particle number density is set to $\rho = 4$ and the total particle number for hydrogels is 256000. Figure~\ref{fig:snapshot} shows a snapshot of the eDPD system, where NPs diffuse inside the interconnected porous network of the thermoresponsive hydrogel.
We consider three different size of hydrogels containing 1536, 2048 and 2560 cross-linked polymer chains with 50 eDPD particles per chain, which correspond to polymer concentrations $C$ = 30 wt\%, 40 wt\% and 50 wt\%, respectively. To study the diffusion of NPs contained in thermoresponsive hydrogels, we introduce spherical NPs with three different sizes of $R = 10$ nm, 12.5 nm and 15 nm. The NPs are constructed from locally frozen eDPD fluid particles, which are nearly uniformly distributed in the spherical NPs~\cite{chen2006flow}. Each NP is treated as an independent rigid body and interacts with solvents, polymers and other NPs governed by the  eDPD equations. The number density of the spherical NPs is also set to 4.

Initially, the polymer chains surrounded by solvent particles are randomly distributed in the simulation region at a low temperature $T_0 = 282$ K. After reaching the thermal equilibrium state~\cite{li2015mesoscale}, 15~NPs with initial temperature $T_0 = 282$ K are randomly placed in the simulation box, and a preprocessing using a short relaxation run is performed to remove the overlap between NPs and polymer solutions. Then, a linear heating method~\cite{li2015mesoscale,tang2016non} is adopted to heat up the eDPD system from $T_0$ to the desired temperature within 2000 reduced time units, or 562 $\upmu$s. Li et al.~\cite{li2015mesoscale} and Tang et al.~\cite{tang2016non} have demonstrated that the linear heating method can effectively control the system temperature while preserving a Gaussian particle temperature distribution. Specifically, each eDPD particle $i$ is coupled with a thermal background of desired temperature $T^B(t)$, and the temperature difference $\Delta T = T^B(t) - T_i(t)$ induces a heat flux
\begin{equation}
    Q_i^B(t) = \lambda C_v \Delta T = \lambda C_v \bigl[ T^B(t) - T_i(t) \bigr],
\end{equation}
where $\lambda$ is a relaxation factor and we set $\lambda = 0.01$ in all our simulations. After heating the eDPD system to the desired temperature, we maintain the temperature for another 562 $\upmu$s to obtain the thermal equilibrium state of the system, and then gather the data for computing statistics.

\section{RESULTS AND DISCUSSION}\label{sec:3}\vspace{-0.15cm}
In this section, we consider the effects of polymer concentration, NP size, and temperature on the diffusion of NPs contained in thermoresponsive hydrogels. To this end, we compute the velocity autocorrelation function (VACF) and mean-squared displacement (MSD) of NPs under different configurations. The VACF and MSD are defined as
\begin{align}
    \mathrm{VACF}(\tau) &= \langle \mathbf{V}(t+\tau) \mathbf{V}^T(t) \rangle, \\
    \mathrm{MSD}(\tau) &= \langle (r(t+\tau)-r(t))^2 \rangle.
\end{align}
Usually, the MSD scales with time lag $\tau$ in the form of a power law, i.e.,  MSD $\propto \tau^{\beta}$. For normal diffusion, $\beta = 1$, while $\beta < 1$ represents the subdiffusion, and $\beta > 1$ corresponds to the superdiffusion~\cite{hofling2013anomalous}. The ensemble average is taken over 20 independent eDPD simulations.

\begin{figure}[t]
\centering
\includegraphics[height = 5cm]{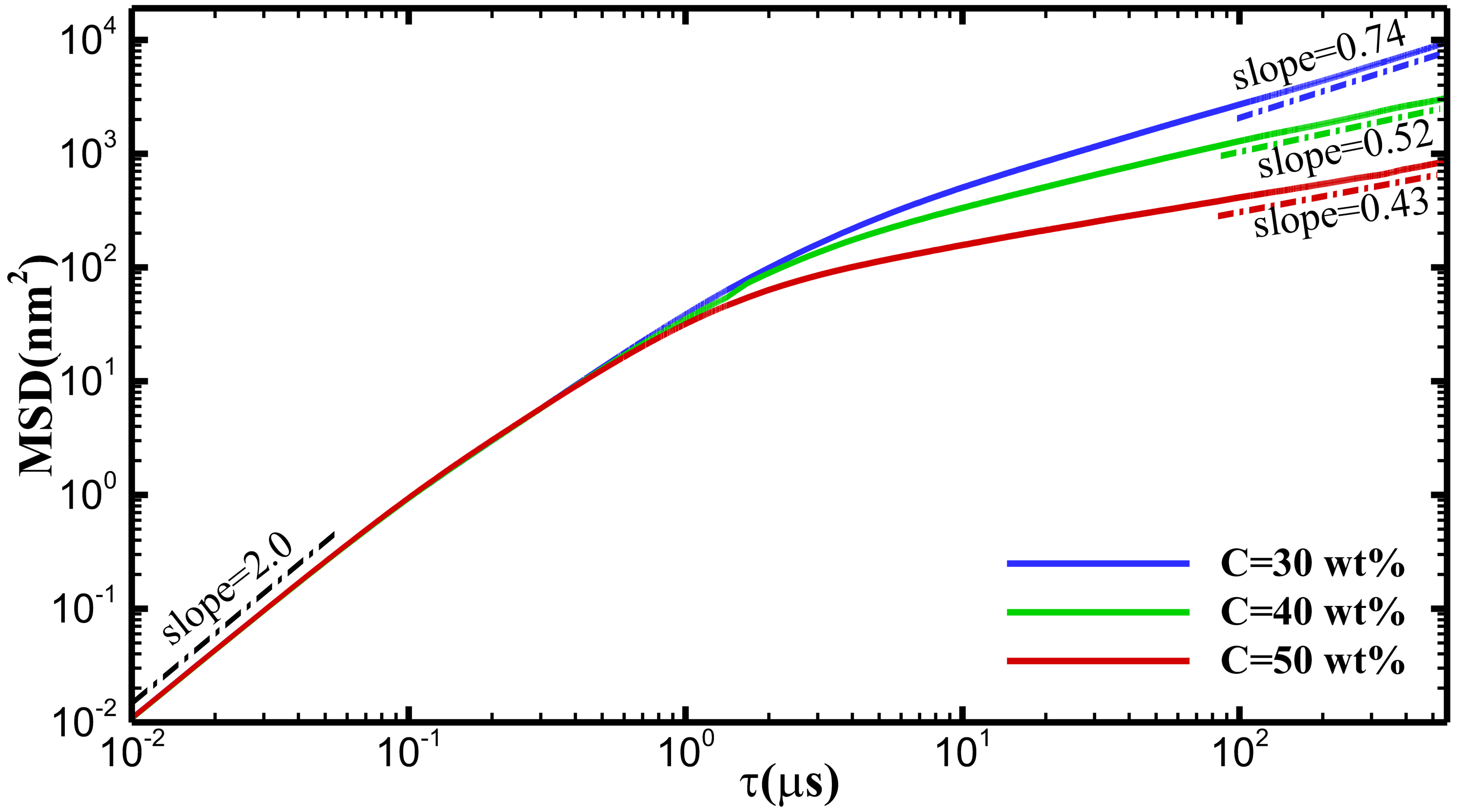}
\caption{Effect of polymer concentration on MSD of NPs with the radius of $R$ = 12.5 nm at $T$ = 282 K.}
\label{fig1:msd_C}
\end{figure}

\begin{figure}[t!]
\centering
\includegraphics[height = 8cm]{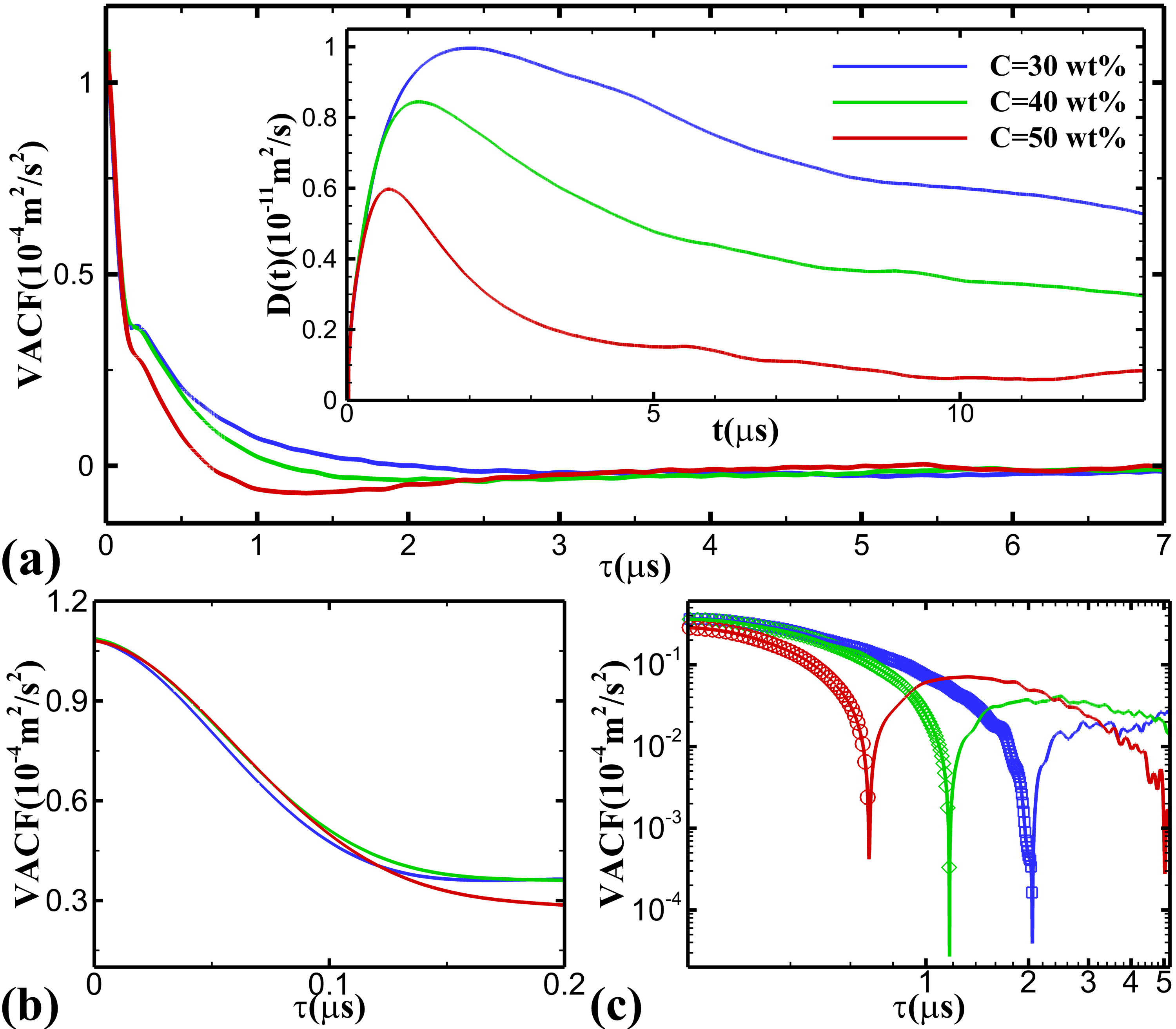}
\caption{Effect of polymer concentration on VACF and diffusion coefficient $D(t)$ of NPs with the radius of $R$ = 12.5 nm at $T$ = 282 K : (a) VACF of NPs in hydrogels with concentrations of $C$ = 30 wt\%, 40 wt\% and 50 wt\% , where the inset shows the diffusion coefficient $D(t)$; (b) zoom-in view of short-time VACF; (c) long-time VACF in log-log scale, where the negative values of VACF are shown using lines without symbols.}
\label{fig2:vacf_C}
\end{figure}

\subsection{Effect of polymer concentration on NP diffusion}
In this subsection, we consider the effect of polymer concentration on NP diffusion and compute the MSD and VACF of NPs contained in thermoresponsive hydrogels with different concentrations. In Figure~\ref{fig1:msd_C} we plot the MSD of NPs with the radius of $R$ = 12.5 nm in hydrogels with concentrations of $C$ = 30 wt\%, 40 wt\% and 50 wt\% at $T$ = 282 K. We observe from Figure~\ref{fig1:msd_C} that for all polymer concentrations, NPs move in a subdiffusive manner after $\tau$ > 10 $\upmu$s, as indicated by the slope. The mean values of the subdiffusive exponent $\beta$ are 0.74, 0.52 and 0.43 for $C$ = 30 wt\%, 40 wt\% and 50 wt\%, respectively. These small $\beta$ values are attributed to the confined motion of NPs in polymer network. As the polymer concentration increases, the MSD and $\beta$ decrease. The reason is that the confinement of the polymer network on NPs increases with increasing polymer concentration. At the initial ballistic regime ($\tau$ < 0.1 $\upmu$s), the MSD is independent of the polymer concentration, because the polymers do not interact with the NPs at short times, thus having little impact on the motion of NPs. The results are in close qualitative agreement with available experimental results~\cite{parrish2018temperature,parrish2017network}.

Correspondingly, we present the VACF and diffusion coefficient of NPs in Figure~\ref{fig2:vacf_C}(a). The initial part of the VACF is shown by a zoom-in plot in Figure~\ref{fig2:vacf_C}(b), while the long-time VACF is displayed in a log-log scale in Figure~\ref{fig2:vacf_C}(c). It can be observed in Figure~\ref{fig2:vacf_C}(c) that the VACF is initially positive and decays exponentially at short times, and then becomes negative because of the cage effect~\cite{fiege2009long}. Because the diffusion coefficient is the integral of VACF, i.e., $D(t) = \frac{1}{3} \int_{0}^{t}\mathrm{VACF}(\tau)\mathrm{d}\tau$, the negative VACF causes the diffusion coefficient $D(t)$ to decrease. From the inset of Figure~\ref{fig2:vacf_C}(a), we can see that $D(t)$ increases with time and reaches a peak at short times, and then decreases before reaching a plateau, which indicates the value of diffusion constant. Additionally, as polymer concentration increases, $D(t)$ decreases. There is little difference in the VACF between different concentrations at short times $\tau$ < 0.1 $\upmu$s (as shown in Figure~\ref{fig2:vacf_C}(b)), but the time of VACF from positive to negative advances with the increase of concentration (as shown in Figure~\ref{fig2:vacf_C}(c)). We note a fluctuating long-time tail of VACF in Figure~\ref{fig2:vacf_C}(c), which is consistent with the experimental results by Grebenkov et al.~\cite{grebenkov2013hydrodynamic}.

\begin{figure}[t]
    \centering
    \includegraphics[width = 0.95\textwidth]{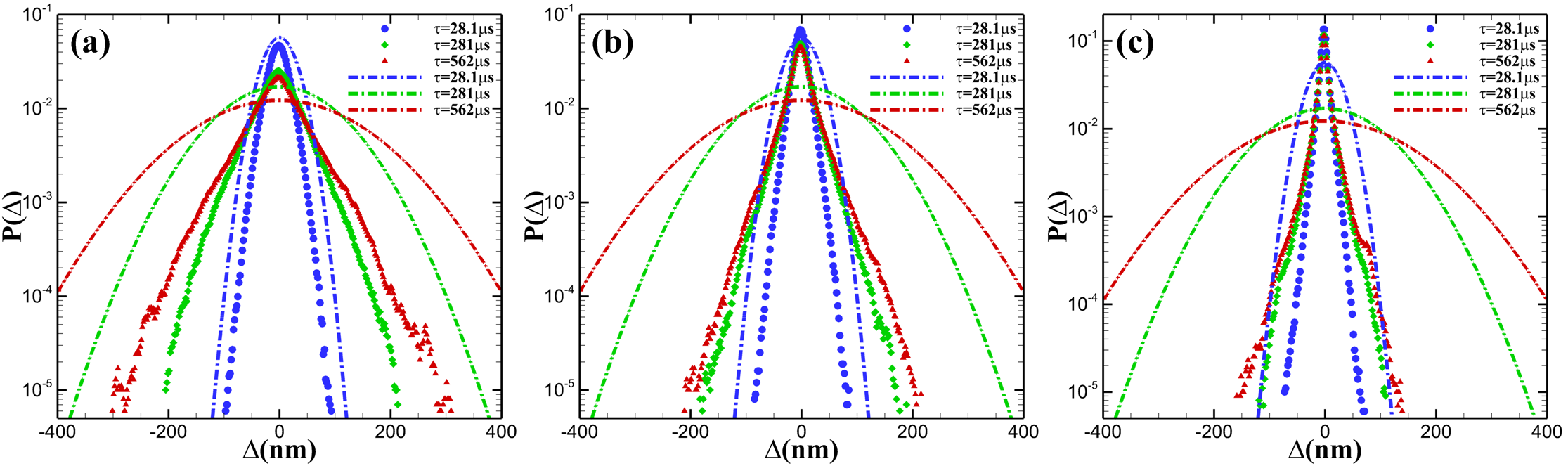}
    \caption{Effect of polymer concentration on van Hove displacement distributions of NPs with the radius of $R$ = 12.5 nm in solvents and hydrogels at $T$ = 282 K : (a) $C$ = 30 wt\%, (b) $C$ = 40 wt\% and (c) $C$ = 50 wt\%. In each figure, solid symbols represent the van Hove displacement distributions of NPs in hydrogels while dash-dot lines stand for the distributions in solvents. The displacement distributions at time interval $\tau$ = 28.1 $\upmu$s, 281 $\upmu$s and 562 $\upmu$s are shown in blue, green and red color, respectively.}
    \label{fig3:vhd_C}
\end{figure}

To quantify the confinement of polymer network on NPs, we introduce the van Hove displacement distributions~\cite{wang2009anomalous,parrish2017network,parrish2018temperature,aufderhorst2012micro}, which characterize the probability distribution of the distance a particle moving along $x$, $y$ or $z$ direction and can be calculated based on~\cite{aufderhorst2012micro}:
\begin{equation}
    \Delta(\tau) = x(t + \tau) - x(t).
\end{equation}
The van Hove displacement distributions of NPs with the radius of $R$ = 12.5 nm in hydrogels with concentrations of $C$ = 30 wt\%, 40 wt\% and 50 wt\% at $T$ = 282 K are shown with symbols in Figure~\ref{fig3:vhd_C}(a),~\ref{fig3:vhd_C}(b) and~\ref{fig3:vhd_C}(c), respectively. As a reference, we plot the van Hove displacement distributions of NPs in solvents under the same condition with dash-dot lines. In each figure, the displacement distributions at time interval $\tau$ = 28.1 $\upmu$s, 281 $\upmu$s and 562 $\upmu$s are shown in blue, green and red color, respectively. From Figure~\ref{fig3:vhd_C}, we can see clearly that the smaller the displacements, the higher the probability of occurrence. In addition, the displacement distribution range of NPs in hydrogels (solid symbols) is smaller than that in solvents (Gaussian distribution, dash-dot lines) at the same time interval. This indicates that the polymer network hinders the motion of NPs. The displacements of NPs in hydrogels with high concentration are time-independent, as noted by the almost overlapping displacement distributions at $\tau$ = 281 $\upmu$s and $\tau$ = 562 $\upmu$s in Figure~\ref{fig3:vhd_C}(c). Moreover, as the polymer concentration increases, the displacement distributions narrow and the largest displacements decrease. At $C$ = 30 wt\%, the largest displacements at $\tau$ = 562 $\upmu$s are greater than 300 nm, while at $C$ = 50 wt\%, the largest displacements at $\tau$ = 562 $\upmu$s are about 150 nm. The NP displacements begin to show time-independent behavior and the spread of the displacement distributions gradually decreases from $\tau$ = 281 $\upmu$s to $\tau$ = 562 $\upmu$s with increasing polymer concentration. Taken together, these results indicate that the NPs are confined in some local regions by the gel network and are subject to stronger hindrance in hydrogels with higher concentrations.

\subsection{Effect of NP size on NP diffusion}
\begin{figure}[t]
\centering
\includegraphics[height = 5cm]{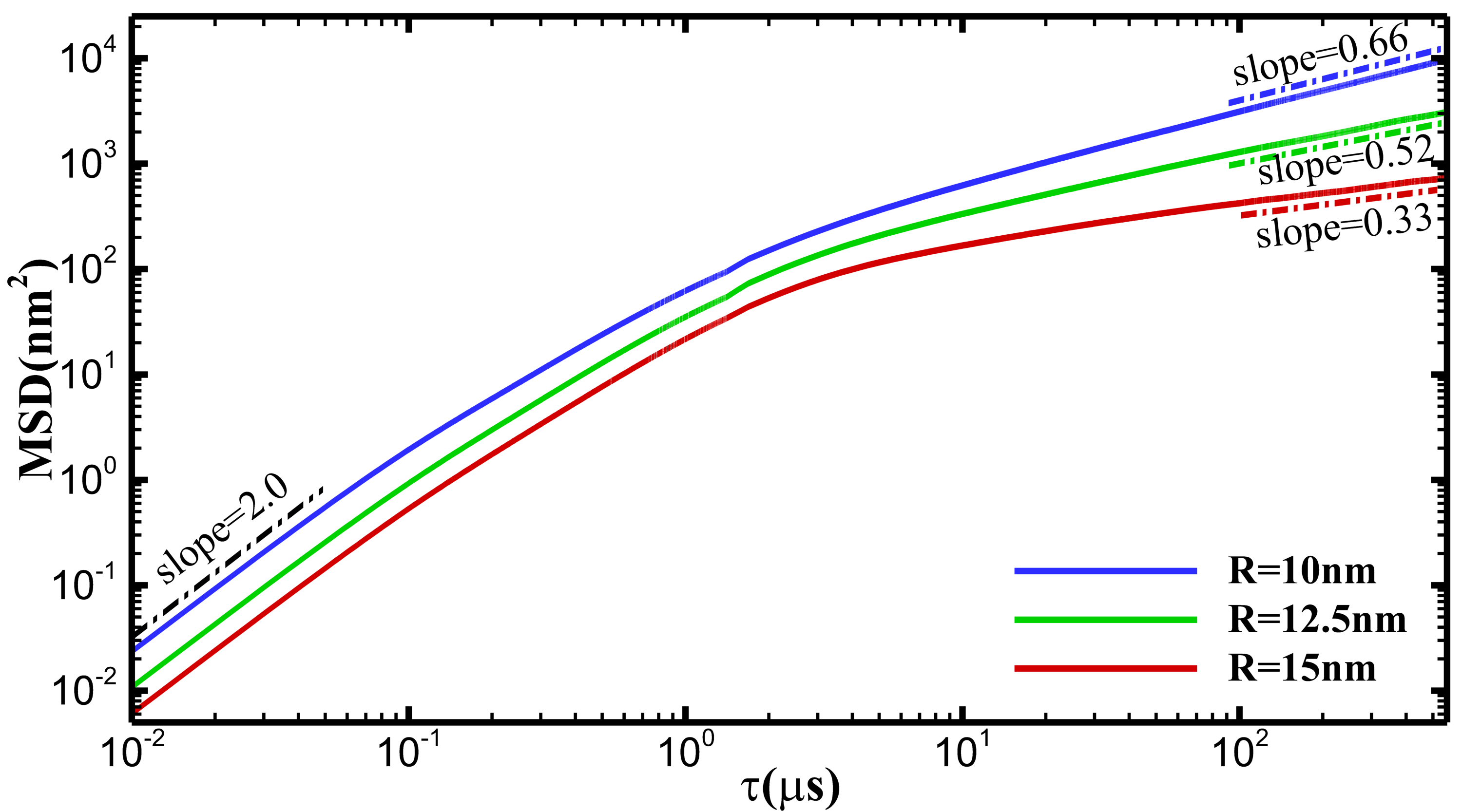}
\caption{Effect of NP size on NPs' MSD in hydrogels with concentration of $C$ = 40 wt\% at $T$ = 282 K.}
\label{fig4:msd_R}
\end{figure}

In this subsection, we consider the effect of NP size on NP diffusion. We present in Figure~\ref{fig4:msd_R} the MSD of NPs with different radius of $R$ = 10 nm, 12.5 nm and 15 nm contained in thermoresponsive hydrogels with concentration of $C$ = 40 wt\% at $T$ = 282 K. We see in Figure~\ref{fig4:msd_R} that the NPs with different sizes move in a subdiffusive manner after experiencing an initial ballistic motion. As the NP size increases, the MSD and subdiffusive exponent $\beta$ decrease. The mean values of $\beta$ are 0.66, 0.52 and 0.33 for $R$ = 10 nm, 12.5 nm and 15 nm, respectively. The main reason is that the space where the NPs can diffuse freely decreases and the NPs are more likely to collide with the polymer network with the increase of NP size.

\begin{figure}[t]
\centering
\includegraphics[height = 8cm]{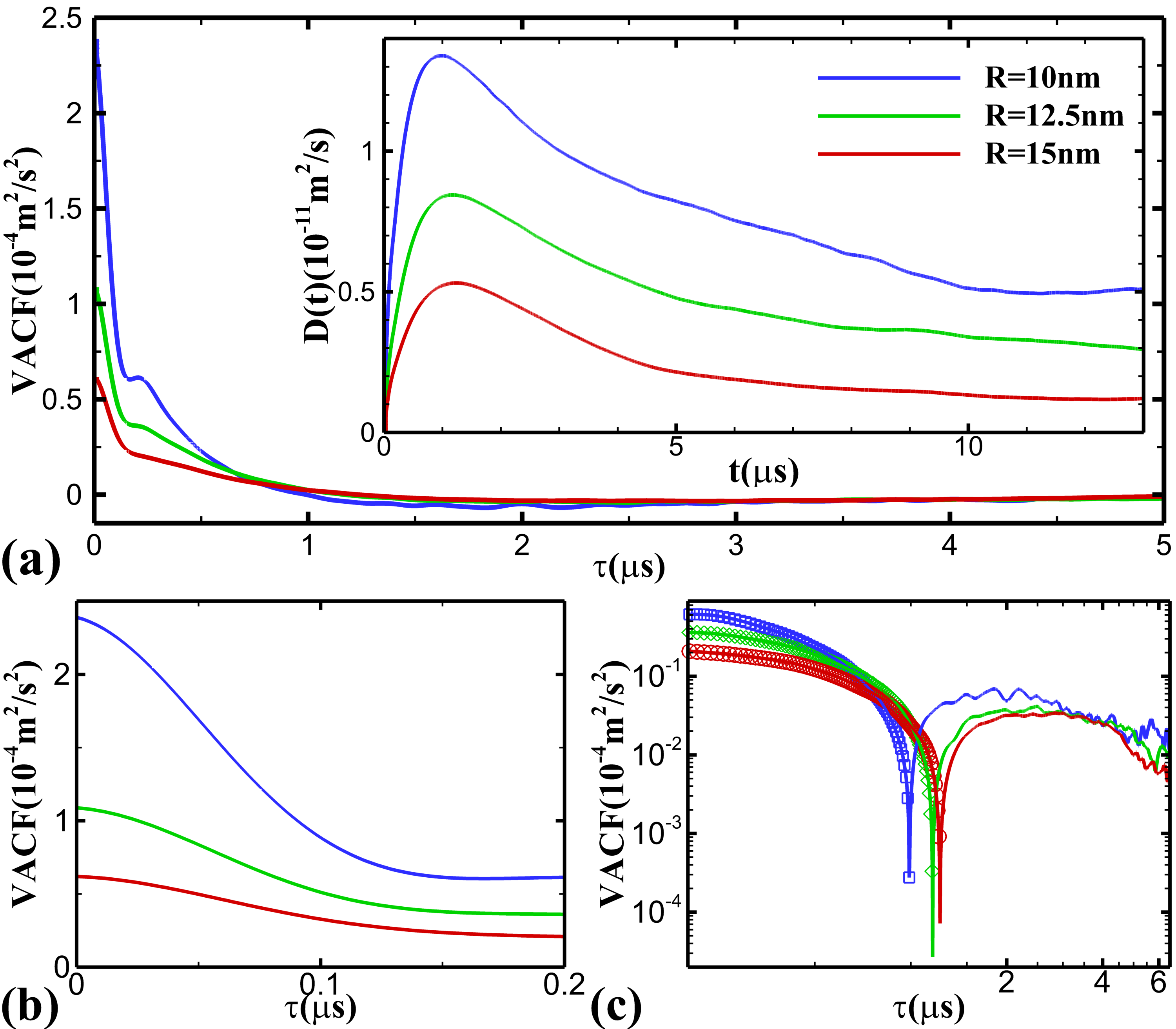}
\caption{Effect of NP size on VACF and diffusion coefficient $D(t)$ of NPs in hydrogels with concentration of $C$ = 40 wt\% at $T$ = 282 K :
(a) VACF of NPs with different sizes of $R$ = 10 nm, 12.5 nm and 15 nm, where the inset shows the diffusion coefficient $D(t)$; (b) zoom-in view of short-time VACF; (c) long-time VACF in log-log scale, where the negative values of VACF are shown using lines without symbols.}
\label{fig5:vacf_R}
\end{figure}

\begin{figure}[t!]
    \centering
    \includegraphics[width = 0.95\textwidth]{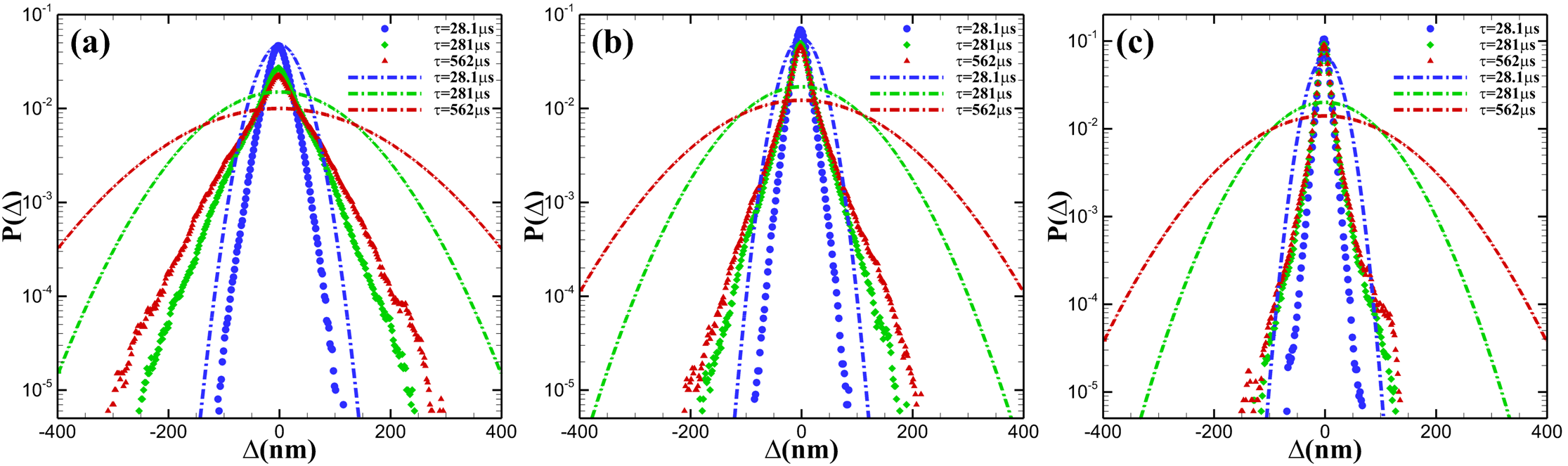}
    \caption{Effect of NP size on van Hove displacement distributions of NPs in solvents and hydrogels with concentration of $C$ = 40 wt\% at $T$ = 282 K : (a) $R$ = 10 nm, (b) $R$ = 12.5 nm and (c) $R$ = 15 nm. In each figure, solid symbols represent the van Hove displacement distributions of NPs in hydrogels while dash-dot lines stand for the distributions in solvents. The displacement distributions at time interval $\tau$ = 28.1 $\upmu$s, 281 $\upmu$s and 562 $\upmu$s are shown in blue, green and red color, respectively.}
    \label{fig6:vhd_R}
\end{figure}

Under the same condition, the VACF and diffusion coefficient $D(t)$ of NPs are shown in Figure~\ref{fig5:vacf_R}(a). The short-time and long-time VACF are given in Figure~\ref{fig5:vacf_R}(b) and~\ref{fig5:vacf_R}(c), respectively. As the NP size increases, the diffusion coefficient $D(t)$ and short-time VACF decrease, as can be seen in Figure~\ref{fig5:vacf_R}(a) and~\ref{fig5:vacf_R}(b). The time of VACF from positive to negative delay with increasing NP size, as shown in Figure ~\ref{fig5:vacf_R}(c).
Moreover, the van Hove displacement distributions of NPs with different sizes in solvents and hydrogels with concentration of $C$ = 40 wt\% at $T$ = 282 K are shown in Figure~\ref{fig6:vhd_R}(a),~\ref{fig6:vhd_R}(b) and~\ref{fig6:vhd_R}(c), respectively. For each NP size, the displacement distributions at time interval $\tau$ = 28.1 $\upmu$s, 281 $\upmu$s and 562 $\upmu$s are shown in blue, green and red color. Similar to Figure~\ref{fig3:vhd_C}, the displacement distribution range of NPs in hydrogels (solid symbols) is smaller than that in solvents (dash-dot lines) in Figure~\ref{fig6:vhd_R}, indicating that the polymer network slows down the diffusion of NPs. In addition, as NP size increases, the van Hove displacement distributions of NPs in both solvents and hydrogels become narrow and the largest displacements at $\tau$ = 562 $\upmu$s decrease. The NP displacements in hydrogels begin to show time-independent behavior with increasing NP size, as can be seen by the decreasing displacement distribution range from $\tau$ = 281 $\upmu$s to $\tau$ = 562 $\upmu$s. The largest NP displacements at $\tau$ = 562 $\upmu$s decrease from about 300 nm to about 150 nm as the NP size increases from $R$ = 10 nm to $R$ = 15 nm. Overall these results indicate that similar to the motion of NPs in solvents (normal diffusion), as NP size increases, the diffusion of NPs in hydrogels decreases, but the latter decreases more than the former because of the confinement of polymer network, which can be seen clearly by comparing the diffusion coefficient $D(t)$ in the inset of Figure~\ref{fig5:vacf_R}(a) with Stokes-Einstein relation.

\subsection{Effect of temperature on NP diffusion}
\begin{figure}[t]
\centering
\includegraphics[height = 5cm]{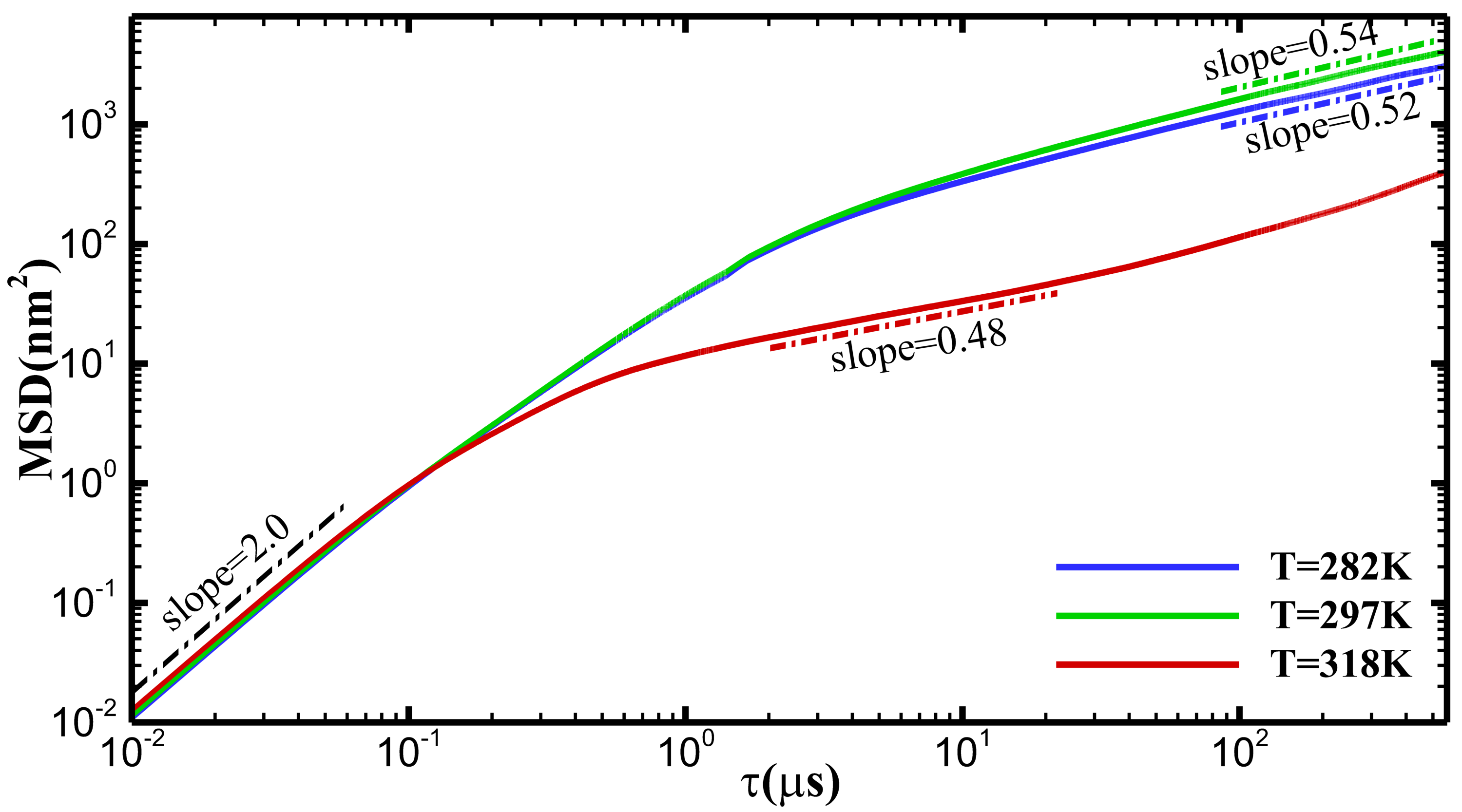}
\caption{Effect of temperature on MSD of NPs with the radius of $R$ = 12.5 nm in hydrogels with concentration of $C$ = 40 wt\%.}
\label{fig7:MSD_T}
\end{figure}

\begin{figure}[t]
\centering
\includegraphics[height = 5cm]{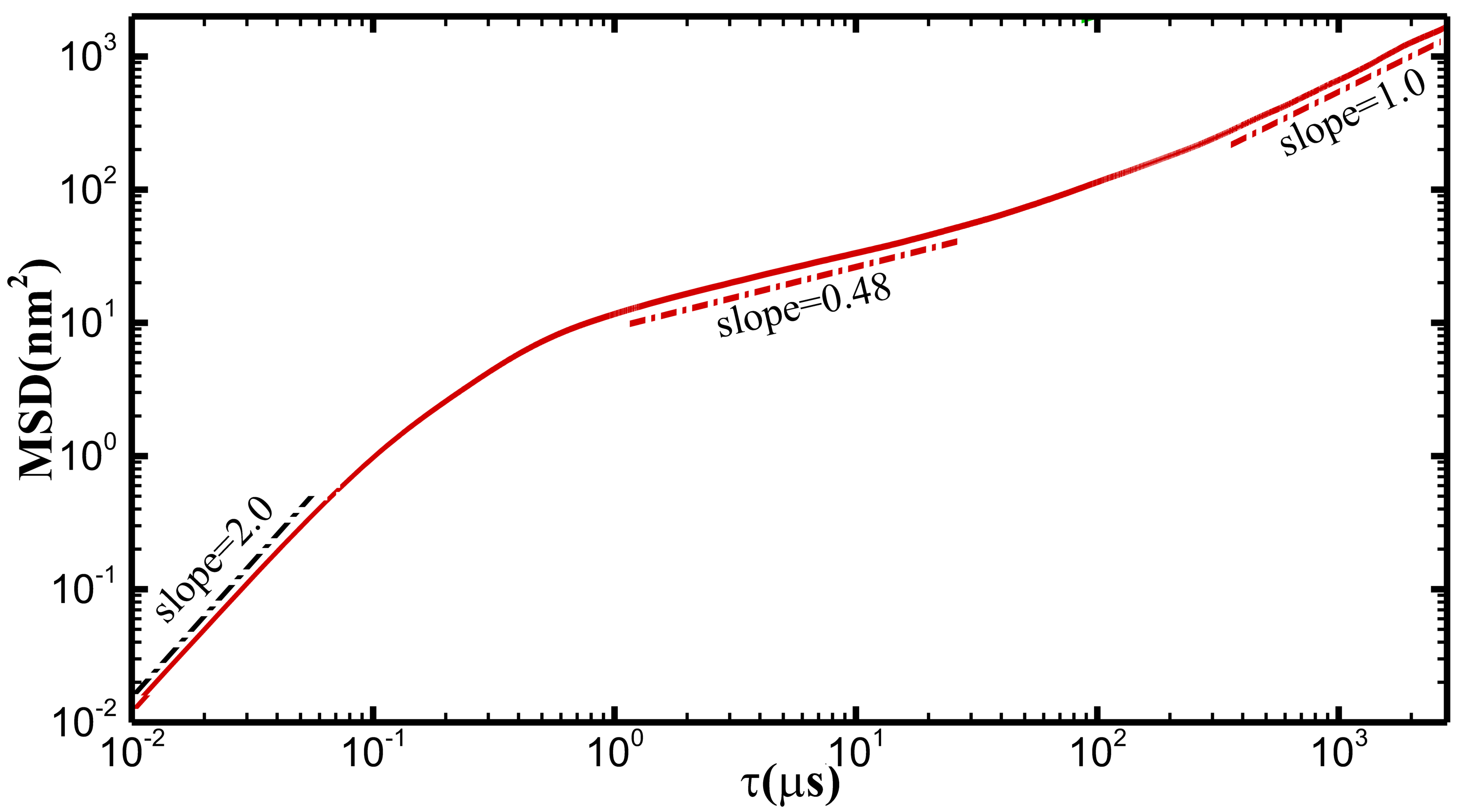}
\caption{MSD of NPs with the radius of $R$ = 12.5 nm in hydrogels with concentration of $C$ = 40 wt\% at $T$ = 318 K.}
\label{fig8:MSD_T_whole}
\end{figure}

In this subsection, we study the effect of temperature on NP diffusion. Figure~\ref{fig7:MSD_T} shows the MSD of NPs with the radius of $R$ = 12.5~nm in hydrogels with concentration of $C$ = 40~wt\% at $T$ = 282~K, 297~K and 318~K. We can see in Figure~\ref{fig7:MSD_T} that for all temperatures tested in this study, the NPs undergo a subdiffusion process after a short-time ballistic motion. As temperature increases and approaches the critical temperature of phase transition ($T_c$ = 300~K), the MSD value and its slope increase. The value of subdiffusive exponent $\beta$ increases from 0.52 at $T$ = 282~K to 0.54 at $T$ = 297~K. With further increase of temperature from 297~K to 318~K, however, the MSD and $\beta$ decrease. If the eDPD simulations are performed for a long time, we expect a final normal diffusion to emerge. As an example, we present in Figure~\ref{fig8:MSD_T_whole} the long-time diffusion process of NPs at $T$ = 318 K. We observe that the NPs initially undergo a short-time ballistic motion (MSD $\propto \tau^2$), followed by an intermediate-time subdiffusive motion from $\tau \approx$ 1 $\upmu$s to $\tau \approx$ 100 $\upmu$s (MSD $\propto \tau^{0.48}$) before a long-time normal diffusion (MSD $\propto \tau$) is recovered.

\begin{figure}[h!]
\centering
\includegraphics[height = 8cm]{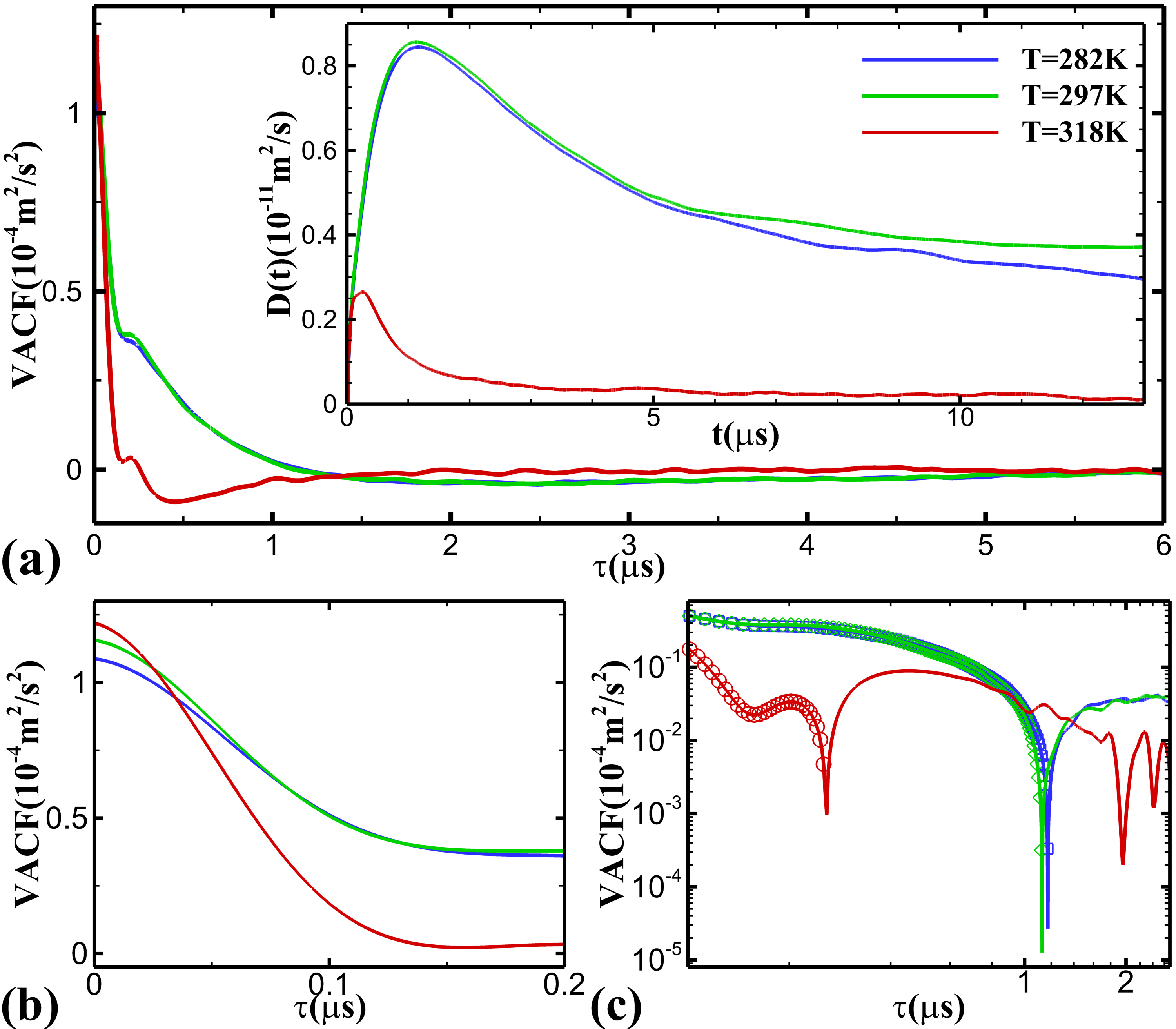}
\caption{Effect of temperature on VACF and diffusion coefficient $D(t)$ of NPs with the radius of $R$ = 12.5 nm in hydrogels with concentration of $C$ = 40 wt\% : (a) VACF of NPs at $T$ = 282 K, 297 K and 318 K, where the inset shows the diffusion coefficient $D(t)$; (b) zoom-in view of short-time VACF; (c) long-time VACF in log-log scale, where the negative values of VACF are shown using lines without symbols.}
\label{fig9:vacf_T}
\end{figure}

The VACF and diffusion coefficient $D(t)$ of NPs are shown in Figure~\ref{fig9:vacf_T}(a), and short-time and long-time VACF are presented in Figure~\ref{fig9:vacf_T}(b) and~\ref{fig9:vacf_T}(c), respectively. From Figure~\ref{fig9:vacf_T}(a), we can see clearly that as temperature increases from 282 K to 297 K, the diffusion coefficient $D(t)$ slightly increases, but it sharply reduces to almost zero with temperature increasing from 297 K to 318 K. Correspondingly, VACF($\tau$ = 0) increases and the time of VACF from positive to negative advances with increasing temperature, as can be seen in Figure~\ref{fig9:vacf_T}(b) and~\ref{fig9:vacf_T}(c).

\begin{figure}[t!]
\centering
\includegraphics[width = 0.9\textwidth]{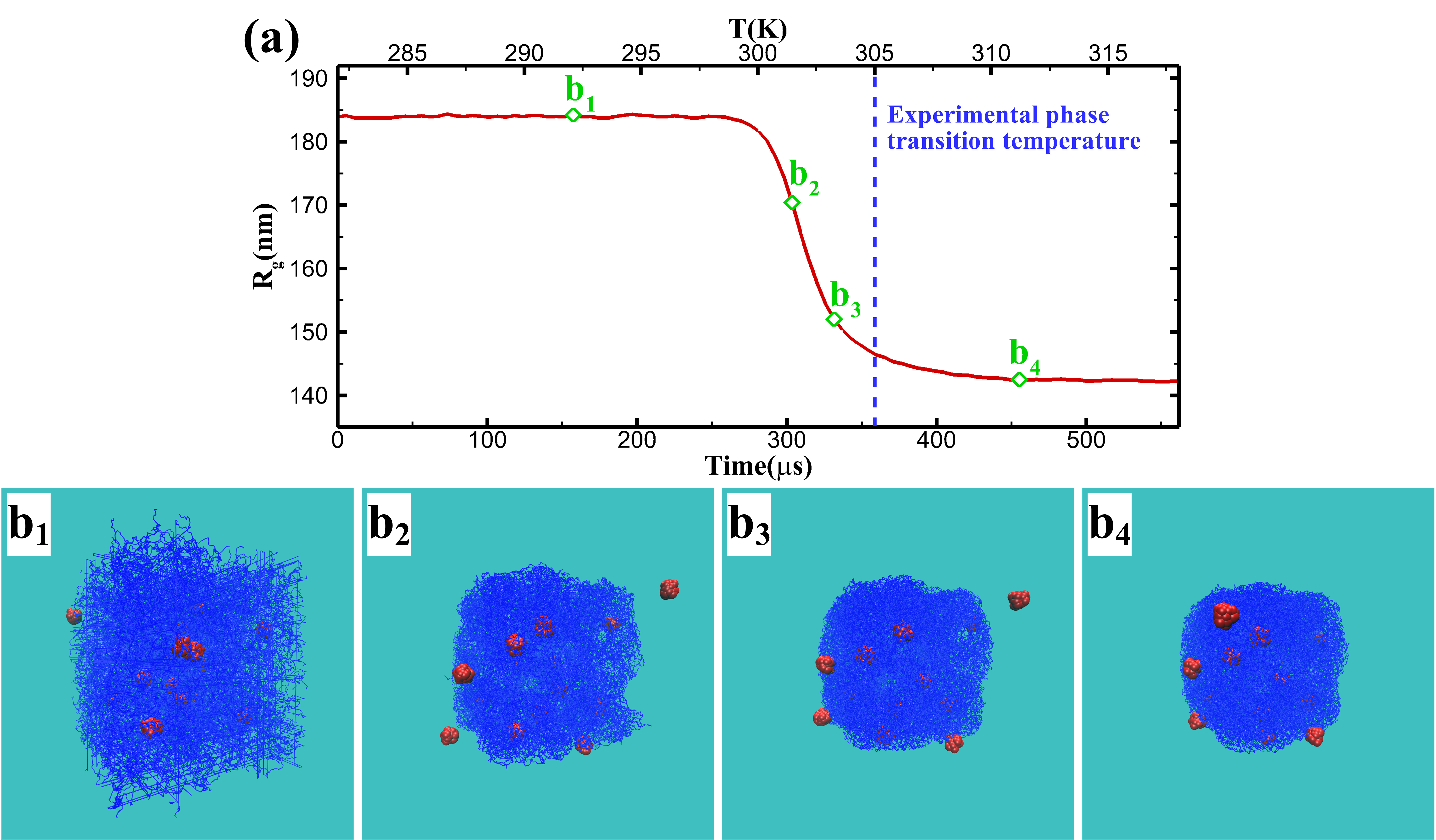}
\caption{(a) Evolution of the radius of gyration $R_g$ of thermoresponsive hydrogels with concentration of $C$ = 40 wt\% during heating from 282 K to 318 K. (b$_1$--b$_4$) show the transient microstructure of NP-hydrogel system corresponding to the changes of $R_g$.} \vspace{-0.1cm}
\label{fig10:Rg_microstr}
\end{figure}

To quantify the configurational change of thermoresponsive hydrogels during the heating process, and analyze its impact on NP diffusion, we compute the instantaneous gyration radius $R_g$. The time evolution of $R_g$ during heating from 282 K to 318 K is shown in Figure~\ref{fig10:Rg_microstr}(a), where a significant decrease of $R_g$ occurs between $T$ = 300 K and $T$ = 305 K, corresponding to a phase transition. The experimental phase transition temperature is also marked in this figure. Specifically, four snapshots of NP-hydrogel system along the $R_g$ curve are presented in Figure~\ref{fig10:Rg_microstr}(b$_1$--b$_4$), and the transient point corresponding to $R_g$ are marked in Figure~\ref{fig10:Rg_microstr}(a). When the temperature is below the critical phase transition temperature, i.e., $T$ < 300 K, the hydrogels are hydrophilic and swollen resulting in a coil state, as shown in Figure~\ref{fig10:Rg_microstr}(b$_1$). At the coil state, the hydrogels have the maximum volume, corresponding to the maximum gyration radius $R_g$. Therefore, the $R_g$ curve has a plateau at $T$ < 300 K, as displayed in Figure~\ref{fig10:Rg_microstr}(a). The confinement of the polymer network on NPs is relatively weak, so the NPs can move further and a larger MSD is observed because of higher kinetic energy, as shown in Figure~\ref{fig7:MSD_T}. However, with the increasing of temperature, the coil-to-globule phase transition emerges near the critical temperature $T \approx$ 300 K, above which the hydrogels become hydrophobic and deswollen, and start to collapse until they turn into compact globule. Some NPs get trapped in the gel network and other NPs escape outside, as shown in Figure~\ref{fig10:Rg_microstr}(b$_4$). The confinement of gel network on encapsulated NPs increases due to the collapse of hydrogels, resulting in the trapped NPs to become localized with decreasing MSD, despite the higher temperature.

\begin{figure}[t!]
    \centering
    \includegraphics[width = 0.95\textwidth]{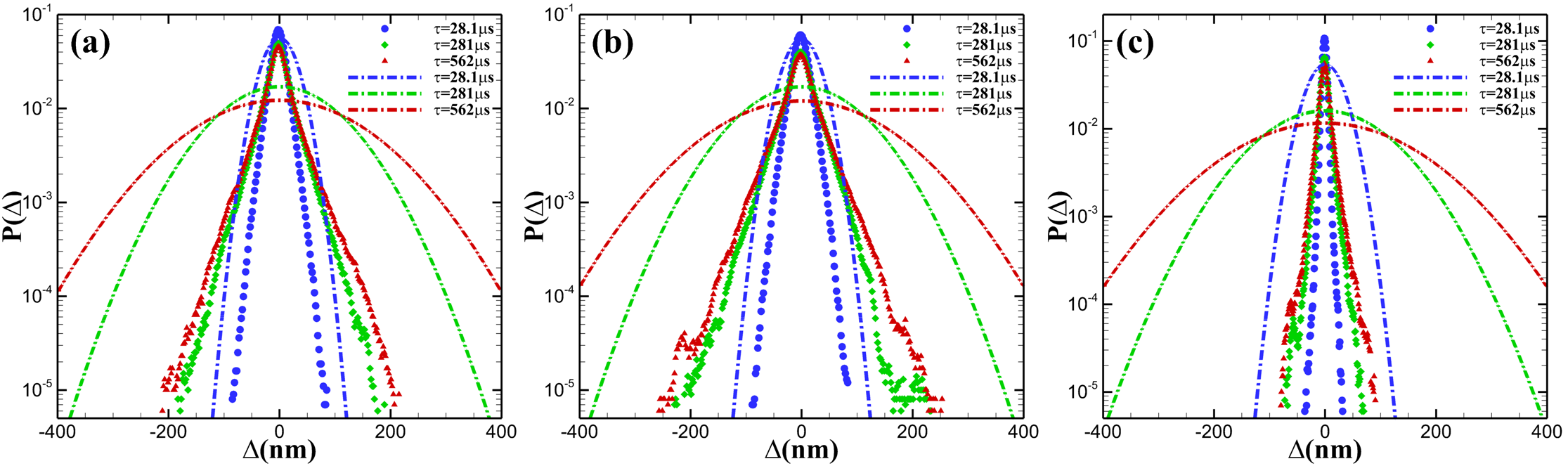}
    \caption{Effect of temperature on van Hove displacement distributions of NPs with the radius of $R$ = 12.5 nm in solvents and hydrogels with concentration of $C$ = 40 wt\% : (a) $T$ = 282 K, (b) $T$ = 297 K and (c) $T$ = 318 K. In each figure, solid symbols represent the van Hove displacement distributions of NPs in hydrogels, and dash-dot lines stand for the distributions in solvents. The distributions at time interval $\tau$ = 28.1 $\upmu$s, 281 $\upmu$s and 562 $\upmu$s are shown in blue, green and red color, respectively.}
    \label{fig11:vhd_T}
\end{figure}

Correspondingly, the van Hove displacement distributions of NPs in solvents and hydrogels at $T$ = 282 K, 297 K and 318 K are shown in Figure~\ref{fig11:vhd_T}(a),~\ref{fig11:vhd_T}(b) and~\ref{fig11:vhd_T}(c), respectively. In each figure, the displacement distributions at time lag $\tau$ = 28.1 $\upmu$s, 281 $\upmu$s and 562 $\upmu$s are shown in blue, green and red color. We see from Figure~\ref{fig11:vhd_T}(a, b) that the displacement distributions of NPs in both hydrogels (solid symbols) and solvents (dash-dot lines) become broad with temperature approaching the critical temperature. At $T$ = 282 K, the largest displacements of NPs in hydrogels at $\tau$ = 562 $\upmu$s are just over 200 nm (shown in red triangle in Figure~\ref{fig11:vhd_T}(a)), while at $T$ = 297 K the largest displacements at $\tau$ = 562 $\upmu$s are greater than 250 nm (shown in red triangle in Figure~\ref{fig11:vhd_T}(b)). As the temperature increases above the critical value, the displacement distributions of NPs in solvents (dash-dot lines) become wider, but the distributions in hydrogels (solid symbols) become significantly narrower, as shown in Figure~\ref{fig11:vhd_T}(c). At $T$ = 318 K, the largest displacements of NPs in hydrogels at $\tau$ = 562 $\upmu$s are less than 100 nm, and all the displacement distributions (solid symbols) are confined in the region bounded by the blue dash-dot line in Figure~\ref{fig11:vhd_T}(c). Additionally, the NP displacements are time-dependent at $T < T_c$, illustrated by the increasing spread from $\tau$ = 281 $\upmu$s (green diamond) to $\tau$ = 562 $\upmu$s (red triangle) in Figure~\ref{fig11:vhd_T}(a, b). However, at $T > T_c$, the displacement distributions become time-independent, shown by the almost overlapping green and red symbols in Figure~\ref{fig11:vhd_T}(c). Simulation results indicate that at temperatures above the critical temperature, the NPs can hardly diffuse (shown in the inset of Figure~\ref{fig9:vacf_T}(a)) and are localized to some small regions, which is attributed to the increase of confinement on NPs by the polymer network due to the collapse of hydrogels. Our simulation results show a close qualitative agreement with the available experimental results~\cite{stempfle2014anomalous,parrish2018temperature}.

\section{CONCLUSIONS}\label{sec:4}\vspace{-0.15cm}
To understand the diffusion process of NPs so as to better control the release of NPs from thermoresponsive hydrogels, we systematically investigated the diffusion of NPs contained in thermoresponsive hydrogels using energy-conserving dissipative particle dynamics (eDPD) simulations. Specifically, we studied the effects of polymer concentrations, NP size, and temperature on NP diffusion, and computed the mean-squared displacement (MSD), velocity autocorrelation function (VACF) and van Hove displacement distributions of NPs.

Our simulation results demonstrate that NPs experience a subdiffusion process after the initial ballistic motion for all polymer concentrations, NP sizes, and temperatures tested in present study. As the polymer concentration or NP size increases, the MSD, subdiffusive exponent and diffusion coefficient of NPs decrease, the van Hove displacement distributions become narrow, and the NP displacements change from time-dependent to time-independent. The VACF of NPs has little difference between different concentrations at short time scales, but the time of VACF from positive to negative advances with the increase of polymer concentration. Similar to the motion of NPs in solvents, the VACF of NPs in hydrogels decreases with increasing NP size, but the diffusion coefficient of NPs in hydrogels decreases more than that in solvents because of the confinement of the polymer network. As temperature increases and approaches the critical temperature of phase transition, the MSD, subdiffusive exponent and diffusion coefficient of NPs all increase. The largest displacements of NPs in hydrogels at $\tau$ = 562 $\upmu$s increase from just over 200 nm at $T$ = 282 K to greater than 250 nm at $T$ = 297 K and the NP displacements are time-dependent. With temperature rising above the critical temperature, however, the MSD and diffusion coefficient of NPs decrease significantly. At $T$ = 318 K, the largest displacements of NPs in hydrogels at $\tau$ = 562 $\upmu$s are less than 100 nm and the NP displacements become independent of time. Moreover, the NPs can hardly diffuse and show fully localized motion at temperatures above the critical temperature, which is attributed to the collapse of hydrogels. Interestingly, despite the increase of 36 K in temperature, the NPs are completely localized at $T > T_c$.

To quantify the configurational change of thermoresponsive hydrogels during the heating process and study the effect of local hydrogel structures on NP diffusion, we computed the instantaneous gyration radius $R_g$ and found a sharp decrease of $R_g$ between $T$ = 300 K and $T$ = 305 K, corresponding to a phase transition. We also showed the long-time diffusion process of NPs in hydrogels at $T$ = 318 K, exhibiting an initial ballistic motion, followed by an intermediate subdiffusion, and a final normal diffusion. We found that the interconnected porous network in smart hydrogels leads to confined subdiffusive motion of NPs, and a variation of the network structure of hydrogels changes the diffusion dynamics of NPs. Consequently, the release of NPs from thermoresponsive hydrogels can be controlled by tuning the hydrogel network characteristics as a response to temperature changes. These computational findings on anomalous NP diffusion in smart hydrogels, which complement available experimental results, offer new and important insights for designing controlled drug release from stimuli-responsive hydrogels, including autonomously switch on/off drug release to respond to the changes of the local environment.

\section*{ACKNOWLEDGEMENTS}\vspace{-0.15cm}
This work was supported by the DOE PhILMs project (No.\ DE-SC0019453)
and by the Army
Research Laboratory (W911NF-12-2-0023).
This research was conducted using computational resources and services at the Center for Computation and Visualization, Brown University.
Y.~Wang would like to thank the China Scholarship Council~(CSC) for the financial support~(No.~201806290181).

\end{document}